# The ethics of biosurveillance

Authors: Devitt, S.K.,[1,2,3] Baxter, P.W.J.[4,5] & Hamilton, G.[4]
Affiliations:
[1] Co-Innovation Group, School of Information Technology and Electrical Engineering, The University of Queensland, Brisbane, QLD 4072, Australia
[2] Defence Science & Technology Group, Fishermans Bend, VIC 3207, Australia
[3] Institute for Future Environments, Queensland University of Technology (QUT), Brisbane, Queensland 4000, Australia
[4] Quantitative Applied Spatial Ecology (QASE) group; School of Earth, Environmental and Biological Sciences, Queensland University of Technology (QUT), Brisbane, Queensland 4000, Australia
[5] School of Biological Sciences, The University of Queensland, Brisbane, Queensland 4072, Australia

Corresponding author:
Dr Kate Devitt
Email: k.devitt@uq.edu.au
Phone: +61 403761076

Kate Devitt ORCID: http://orcid.org/0000-0002-6075-4969
Grant Hamilton ORCID: https://orcid.org/0000-0001-8445-0575
Peter Baxter ORCID: https://orcid.org/0000-0003-0585-9482

## 1. Abstract

Governments must keep agricultural systems free of pests that threaten agricultural production and international trade. Biosecurity surveillance already makes use of a wide range of technologies, such as insect traps and lures, geographic information systems, and diagnostic biochemical tests. The rise of cheap and usable surveillance technologies such as Remotely Piloted Air Systems (RPAS) presents value conflicts not addressed in international biosurveillance guidelines. The costs of keeping agriculture pest-free include privacy violations and reduced autonomy for farmers. We argue that physical and digital privacy in the age of ubiquitous aerial and ground surveillance is a natural right to allow people to function freely on their land. Surveillance methods must be co-created and justified through using ethically defensible processes such as discourse theory, value-centred design and responsible innovation to forge a cooperative social contract between diverse stakeholders. We propose an ethical framework for biosurveillance activities that balances the collective benefits for food security with individual privacy: 1) Establish the boundaries of a biosurveillance social contract; 2) Justify surveillance operations for the farmers, researchers, industry, the public and regulators; 3) Give decision makers a reasonable measure of control over their personal and agricultural data; and 4) Choose surveillance methodologies that give the appropriate information. The benefits of incorporating an ethical framework for responsible biosurveillance innovation include increased participation and accumulated trust over time. Long term trust and cooperation will support food security, producing higher quality data overall and mitigating against anticipated information gaps that may emerge due to disrespecting landholder rights.

## 2. Introduction

The need to detect plant pests and diseases has increased in modern times. Natural and anthropogenic factors have led to both a dramatic increase of spread into new areas (Levine and D'Antonio, 2003) and the evolution of strains overcoming current control techniques (Zhan et al., 2014). Food production practices will need to change to feed a growing global population and to offset crop losses from increasing incursions of pests and pathogens (Griffin, 2014, p. 694; Mills et al., 2011; Strange and Scott, 2005). Modern biosecurity surveillance involves increasingly sophisticated pest and disease detection, data collection and analysis. There is a growing need to assess and adopt technologies that will maintain the critical role of biosecurity surveillance in the face of increasing threats to protect global food security.

The opportunities of autonomous biosurveillance are to identify pest and disease incursions faster, with greater precision, reducing the cost and time required to successfully intervene. Autonomous biosurveillance integrated into digital farming practices can help reduce the number of crop sprays and increase their precision (Gebbers and Adamchuk, 2010),[1] reducing the risk to the farmer and farming community and improving farming sustainability. Reducing input costs and increasing production through digital agriculture systems can lead to productivity gains of 10-15% in cropping systems (Keogh and Henry, 2016). Additionally, greater data and digital integration in agriculture may lead to improved decision-making, increased efficiency, economic and sustainability gains (Wolfert et al., 2017; Sonka, 2016). Better biosurveillance enhances States' capabilities to respect, protect and fulfil human entitlements to food (United Nations Committee on Economic Social and Cultural Rights; Johnson, 2015) for sustainable agricultural production (Altieri, 2018) and to protect international trade (Simpson and Srinivasan, 2014).

But a serious cost of keeping agriculture pest-free using modern technologies may be an invasion of privacy (United Nations General Assembly). Risks of autonomous biosurveillance arise when there is ubiquitous and indiscriminate data capture of farming lands and likely misuse of this information by relatively unregulated 3rd parties (Zuboff, 2019)[2]. Governments and activists can use land use data for political reasons (Pearce, 2018, July 25; Phelps, 2018, November 7; Daly et al., 2019; Coote, 2018, August 28; Davies, 2019, January 31) and neighbours and community members might infringe on farmer privacy by operating drones over property lines. Biosurveillance technology providers may practise surveillance capitalism: selling data to third parties to manipulate human behaviours for business ends, rather than protecting and supporting farmers and communities (Zuboff, 2019). Beyond privacy issues there are also considerations of distributive justice for ethical biosurveillance. However, a thorough examination of these issues are out of scope for this paper.

In sum, advances in surveillance technology may threaten freedom of expression and association; inhibiting the functioning of a free and just society. There is a conflict between the implementation of biosecurity surveillance for food security and the rights of individuals in farming communities not to be surveilled or forced into unethical data relationships. Innovation in surveillance technologies must be developed and applied responsibly to manage the conflict between the right to food and the right to privacy.

This paper investigates the impact of innovation in biosurveillance on people. The paper provides detail on what biosurveillance is and how human rights are at risk as technologies radically increase their data capture. The paper includes a case study on Australian biosecurity and autonomous biosurveillance technologies to provide a real-world example of a high-technology agricultural environment. We go on to propose an ethical framework for biosurveillance activities that balances the collective benefits for food security with individual privacy:

1. Establish the boundaries of a biosurveillance social contract;

---

[1] Technologies such as drone-mounted sprays target specific plants, and even specific parts of plants.
[2] It is in the commercial interest of data capture/analysis technology providers to collect and monetise data about many types of farming activities.

2. Justify surveillance operations for the farmers, researchers, industry, the public and regulators;
3. Give decision makers a reasonable measure of control over their personal and agricultural data; and
4. Choose surveillance methodologies that give the appropriate information.

## 2.1. Surveillance on farm

The increase in global trade and biosecurity threats has taken place alongside rapid advances, and cost-efficiency, in surveillance technology and data volume. Given its multiple goals––from early warning systems, through evaluation and containment of the range of an invasion, to proof of freedom (confirmation of areas being pest-free for trade purposes)––biosecurity surveillance already makes use of a wide range of technologies, such as insect traps and lures, geographic information systems, and diagnostic biochemical tests ([Kalaris et al., 2014](#)). However, almost all components of biosurveillance, even long-established methods, are now subject to technological reinvention. For example, there are clear benefits for establishing regional networks of wireless-communicating early-warning "smart traps" ([McIntyre, 2016](#); [Kimber et al., 2016](#)) especially in remote areas[3]. Physical sampling—whether on-ground field sampling, or strategic sampling at marketplaces (similar to quality-control)—may become increasingly automated with robots displacing or augmenting human operators. So, while traditional farm biosurveillance involves periodic inspection and sampling of crops, produce and other biological materials, new technologies such as satellite imagery and RPAS (Remotely Piloted Air Systems, formerly UAVs or drones) are increasingly employed to scan the landscape for pests. These technologies can lead to more comprehensive surveillance to ensure rapid detection of incursions and faster responses, but also significantly increase the volume of data collected from private land. Such data can reveal personal or commercial information, including potentially sensitive material, for example the movements of individuals or crop husbandry approaches, that may impact on a farmer's freedom to operate; to do what they do best with minimal outside interference based on society's trust in farmers ([Lush, 2018](#)). Farmers have reasons to be wary of sharing farm information as doing so poses several potential risks ([Yiridoe, 2000](#)). Thus, while enhanced biosurveillance offers great promise to global food security, there is a clear and urgent need to consider ethical issues for individuals – such as violations of privacy, unauthorized information use or access, and loss of control over personal information – that emerge in parallel with technical innovation. Here we present ethical considerations to guide the anticipated future and furore of biosurveillance value and rights conflicts to guide responsible innovation.

## 2.2. Rights conflicts

Rights conflicts are common ([Kamm, 2007, pp. 262-301](#)) and there are conceptual frameworks to manage them. Rights theorists usually agree that there is no absolute methodology for evaluating right prioritisation, instead each instance of rights conflict needs to be considered within the context of its occurrence. Thus, the conflict between surveillance and privacy in an agricultural context must be distinguished from similar conflicts in other domains (such as city-wide crime surveillance) and thoroughly examined across multiple viewpoints and rights-holders to form an ethical solution.

---

[3] Smart technologies include insect capture based on GPS or weather parameters; in-field diagnostics (for example, loop mediated isothermal amplification (LAMP) assays); biosensors; and wireless or telemetry data transmission from the field (McIntyre, 2014).

Some have argued that when a conflict is properly analysed and understood, it will resolve into a composible set of actions based on a holistic assessment of the context ([Steiner, 1996](#))—a view known as *specificationism* ([Steiner, 1996](#); [Kamm, 2007](#)). Specificationism supposes that rights never actually conflict once the full context of their involvement is understood. Critics have argued that specificationists require an impossible level of epistemic investment to map out rights in order to know how a conflict should play out ([Thomson, 1990](#)). Also, even when biosurveillance can be justified, some would argue that privacy violations will still occur as a conflict and thus need to be acknowledged. Finally, though individual farmers may fight to prioritise privacy, society is increasingly demanding transparency and control around farming practices ([Arnot, 2018](#)). So, the *right to farm* must compete against other rights and obligations in an evolving society. Arnot ([2018](#)) argues that economic and social pressures will force farmers to report on their practices and suffer the consequences if consumers reject their methods regardless of their right to farm. For example, the social outrage from gruesome video footage filmed on live animal export ships is likely to lead to the shutting down of the $1.8 billion Australian live export industry ([Burton et al., 2018](#)). Thus, it behooves farmers to develop trust with consumers and government organisations through establishing an agreed level of transparency for their practices. The *right to farm* may be better reframed as the *right to farm sustainably and transparently*.

We believe that a great deal can be achieved for diverse stakeholders through an ethically focused investigative process to audit and benchmark the scope of privacy rights violation risks versus duties to be transparent and farm sustainably. We advocate for evaluative feedback processes to be built into biosurveillance regulatory frameworks to ensure that epistemic gaps in rights and duties can be corrected over time.

In this paper we argue that while food security trumps *absolute* privacy for individual farmers, a *scientific* approach to *limited* data surveillance based on sophisticated predictive models can reconcile respect for privacy with sustainable agricultural practices. These models will allow patchy data to provide sufficient insight to improve pest and disease decision making without unnecessary invasion of individual activities—some of which have nothing to do with farming itself. As an analogy, an individual's privacy may be trumped by society's duty to investigate the violation of some other right, e.g. *some* of an individual *p*'s right to privacy may be trumped during a police inquiry into *p* committing credit card fraud, such as police having a right to access *p*'s bank records. Similarly, *some* farming privacy is reasonable to forfeit under particular biosecurity circumstances, such that the government is justified setting up more stringent data collection on a specific property for a set duration. The specifics of these contexts need clarification to ensure that governments do not overstate their right to surveil. We take the current population, economic and climate crisis to be unprecedented circumstances where individual privacy can be less morally urgent than the moral imperative to gather appropriate data to improve the likelihood of sustainable human societies ([Christensen et al., 2018](#)).

## 2.3. An ethical framework

We believe that biosurveillance policies require social justice considerations across property, state and national boundaries whilst respecting the dignity of the individual. Privacy on one's own land should be defended as a natural right to protect individual physical freedom. Such a natural right has not been discussed historically in part because ubiquitous and comprehensive surveillance has been unlikely or impossible. Transforming physical privacy into a natural right is in line with Bernal's (2014) argument that internet privacy allows people to function freely on the internet—which is central to daily living, thus should be considered more akin to a natural right than a legal right. We believe physical privacy in the age of unprecedented surveillance ought to be treated more like a natural right, as it exists independently of the culture or laws of individual states.

Therefore, we argue that *physical and digital privacy in the age of ubiquitous aerial and ground surveillance is a natural right to allow people to function freely on their land. Privacy can be infringed if and only if surveillance methods are co-created and justified through both scientific and social agreement for sustainability*.

Rather than resolving the issue with a right's framework, we argue that an effective ethical framework for biosurveillance is John Rawls' (1971/1999) theory of justice, which seeks to ensure fair cooperation of free and equal stake-holders. Rawls imagined the fairest social contract is one negotiated and agreed to by each individual in society, irrespective of their socioeconomic circumstances or capacities, and best realized via universalized, top-down governance using a range of mechanisms such as policy (e.g. regulating biosurveillance), infrastructure (e.g. internet connectivity, transportation routes), service provision (e.g. data hosting, data analysis and incursion guidance), guidelines (e.g. integrated pest management) and regulation (e.g. restrictions on pesticide applications). Modern biosurveillance requires national government oversight to maintain international trade agreements (Food and Agriculture Organisation of the United Nations, 2011). Therefore, we adopt a top-down federal regulatory approach in this paper, while acknowledging the potential for communitarian arguments against this.

We propose the following recommendations for ethical biosurveillance:

1. *Establish the boundaries* of a biosurveillance social contract;
2. *Justify surveillance operations* for the farmers, researchers, industry, the public and regulators;
3. Give decision makers a *reasonable measure of control* over their personal and agricultural data; and
4. *Choose surveillance methodologies* that give the appropriate information.

In order to justify these ethical recommendations, it is important to understand biosecurity and autonomous biosurveillance. In the following sections we provide background from which analysis of the framework can proceed. We have chosen two case studies to illustrate contemporary approaches to biosecurity and biosurveillance in a modern technologically advanced, democratic agricultural nation, Australia. We will consider each recommendation in within this context.

## 3. Biosecurity

Biosecurity is the set of socio-scientific practices that limit the risk of pest and disease outbreaks on- and off-farm through physical, chemical, ecological and strategic actions. Biosecurity actions include segregation, containment, quarantine, surveillance, monitoring, inspection, and isolation ([Ingram, 2009](); [Maye et al., 2012]()). The spread of more than 1,000 plant pathogens is now regulated by the International Plant Protection Convention (IPPC) ([Food and Agriculture Organisation of the United Nations, 2018]()). There is an equivalent organisation in animal health, the Office International des Epizooties (OIE) ([World Organisation for Animal Health, 2018]()). Both of these are complimented by the WTO Agreement on the Application of Sanitary and Phytosanitary Measures (SPS Agreement)([World Trade Organisation, 2018]()). The WTO agreement requires that any measures justified on biosecurity grounds do not impose unnecessary, arbitrary, unjustifiable, or disguised restrictions on trade. The IPPC's international standard ISPM 6 *Guidelines on Surveillance* state that, "national plant protection organisations (NPPOs) should be in a position to validate declarations of the absence or limited distribution of quarantine pests" ([Secretariat of the International Plant Protection Convention (IPPC), 2016-4]()). So biosurveillance is part of a coordinated, global program of biosecurity to allow trade, yet also protect agricultural production.

Biosecurity governance has not only scientific but also political and cultural aspects as nations reinforce state sovereignty through biosecurity legislation ([Maye et al., 2012]()). Thus, government constructions of biological threat may serve protectionist agendas against the liberal trade endorsed by the world trade organisation. Braun ([2007]()) argues that biosecurity weds 'geopolitics' (governing domestic and international security) with 'biopolitics' (regulation of populations). The political dimension of biosecurity is particularly relevant in the context of interrogating the justification of biosurveillance activities and privacy implications. In particular, the perception and evaluation of risk and hazard is subjective and culturally determined making data collection vulnerable to political decisions over scientific ones.

The account below is a case study of the value of biosecurity and recommended practices in Australia that may overlap with some values and practices in other countries, though specifics will vary. The choice of Australia is pertinent to the arguments in this paper as a tension exists in Australia's scientific justifications for biosecurity measures and its political and economic agenda.

### 3.1.  CASE STUDY: Australia

Australia's geographic isolation and reputation as a high-quality food producer makes it strategic to implement strong biosecurity measures including strict quarantine measures ([Bashford, 1998]()).  Average farm profits would be $12,000 to $17,500 lower without Australia's current biosecurity system ([Commonwealth of Australia, 2015, p. 123]()).  For consumers, 25% of food product costs are due to pest, diseases and weeds ([CSIRO, 2009]()). To meet biosecurity measures, individual farmers are expected to keep records of all farm inputs (and outputs) so that actions can be traced back or traced forward in the event of a pest incursion or disease outbreak ([Animal Health Australia and Plant Health Australia, 2013]()).

From a trade perspective, Australia has relied on a policy of `competitive productivism' ([Dibden et al., 2009](#)), whereby agriculture is unsubsidised and highly productive in order to win markets[4] Australia's openness to free trade is somewhat at odds with their strict restrictions of risky imports that has been classified by the WTO as 'protectionist' and 'trade limiting' ([Dibden et al., 2011](#)). Indeed, the use of quarantine provisions has been more conservative and risk averse for products where the domestic industry is strong ([Beale, 2008](#)).
To grow international trade and improve access premium markets, the Australian Government has invested $200 million in biosecurity surveillance and analysis as part of a larger goal to build a more resilient, productive, profitable and more sustainable Australian agriculture sector ([Commonwealth of Australia, 2015](#)). The Australian government sees it as vital to move from paper-based traceability to digital traceability to maintain market confidence in exports and to efficiently respond to incidents. Digital traceability requires investment in information systems to "better support enhanced surveillance and analysis and the implementation of the Biosecurity Act, 2015" ([Commonwealth of Australia, 2015, p. 118](#)). Traceability information systems need firm ethical guidance during their design to ensure farmer privacy despite information demands arising from national security and global markets. Biosecurity tracking on farm can include records of physical, chemical, ecological, and strategic actions.

### 3.1.1. Physical biosecurity

Physical biosecurity barriers include restricting animal and/or plant materials from entering or leaving the border, a region or specific property, via quarantine. It also means documenting and keeping track of human movements on farm and between farms (such as staff who work across different farms). Farm visitors in Australia are urged to respect farm biosecurity by calling ahead or visiting the office before entering a property—see *Farm Biosecurity sign* for Australian farmers—*Figure 1*.

---

[4] Compare and contrast this approach with the UK that protects marginal agriculture with trade barriers and subsidies Braun, B. (2007). Biopolitics and the molecularization of life. *cultural geographies*, 14(1), 6-28. http://journals.sagepub.com/doi/pdf/10.1177/1474474007072817.

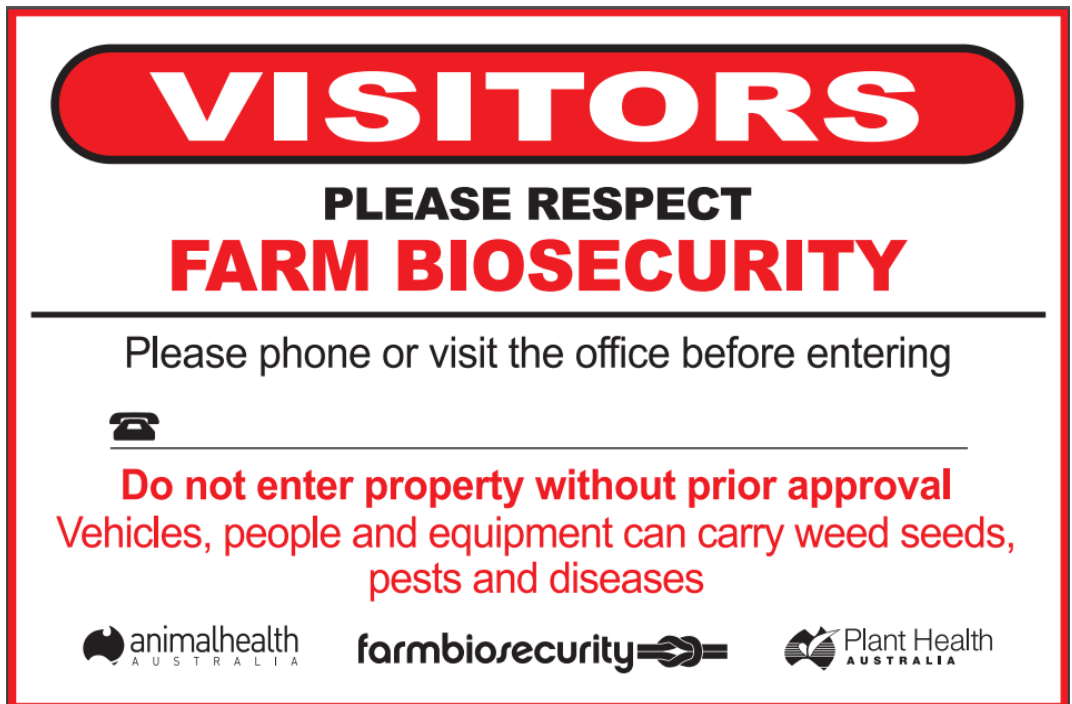

*Figure 1.* Farm Biosecurity sign for Australian farmers (Animal Health Australia and Plant Health Australia, 2013). Used with permission from Plant Health Australia

Trucks and farm equipment must obey 'come clean, go clean' procedures including full wash downs upon arrival to properties and when leaving to reduce the spread of weeds or materials that can harbour pests (Department of Agriculture Fisheries & Forestry, 2013). Good biosecurity farm management practices include processes tailored for each industry, e.g. see the biosecurity module of 'MyBMP' best management practice guide for Australian cotton (Cotton Research Development Corporation and Australian Government National Landcare Programme, 2018). Good management requires communication of biosecurity protocols to all farm personnel, consultants, contractors, visitors and the creation, maintenance and use of a farm biosecurity plan. When done well, biosurveillance can lead to smarter physical interventions, but at the same time individuals must agree to their data being used for potentially restricting their own freedoms—i.e. if data used for quarantine it may be more targeted but could lead to stricter limits on movement and behaviour.

### 3.1.2. Chemical biosecurity
Chemical biosecurity includes use of pesticides, herbicides and fumigation treatments to reduce weeds, pests and diseases. Chemical applications are expensive, time-consuming and potentially toxic to humans, beneficial insects and the wider environment and in Australia require certification and precise adherence to regulations for use (ChemCert, 2018). Good biosurveillance should facilitate 'smarter' applications of chemical treatments leading to efficiency and lower risk of pollution and environmental damage.

### 3.1.3. Ecological biosecurity
Ecological interventions incorporating agroecology (Altieri, 2018) require broad knowledge of complex ecological systems and short and long-term consequences of other biosecurity interventions. Agroecological methods strengthen an agricultural enterprise's 'immune system' by enhancing functional biodiversity by encouraging the natural enemies of pests, allelopathy and antagonists through the creation of appropriate habitats and through adaptive management ([Pimbert, 2018, p. 10](); [Altieri, 2018]()). Enhancing beneficial biological interactions and synergies promotes sustainable production and resilience.

### 3.1.4. Strategic biosecurity

Strategic biosecurity barriers are actions that minimise the likelihood of pest or disease on farm by optimising best practice guidelines for individual farms and on-farm decision-makers. Strategic measures include rigorous on-farm hygiene, and diligence in purchasing seeds, fertilizer and animal feed that is fit for purpose with minimal weed seeds and used within its expiry date. Animal feed is stored clean and dry. Ruminants are not fed animal or bone meal to reduce threat of bovine spongiform encephalopathy (BSE or Mad Cow disease), and pigs are not be fed swill (material of mammal origin such as milk) to reduce the spread of viruses extremely harmful to pigs, other livestock and humans[5]. Water supplies and storage on farm are kept clean and uncontaminated because diseases, pests and weed seeds are easily distributed by flowing water.

### 3.1.5. Conclusion

Australia provides a case study in biosecurity due to its reputation as a high-quality producer and its relative isolation from international markets. It is in Australia's interests to invest in the most effective biosecurity measures available and to encourage farmers to incorporate physical, chemical, ecological, and strategic biosecurity measures in their day-to-day farming practice. The next section discusses the specific tasks of biosurveillance within the aims of biosecurity.

## 4. Biosurveillance

Biosurveillance is the act of observing biological plant and animal systems for evidence of pests and diseases to determine methods to eradicate the immediate threat and put measures in place to prevent the further spread of the contaminants[6]. The United Nations Secretariat of the International Plant Protection Convention (IPPC) *Surveillance Guidelines* '1.2 Collection, storage and retrieval of information' ([2016]()) state that National Plant Protection Organisations (NPPOs) ought to keep:
- a national repository for plant pest records
- a record-keeping and retrieval system,
- data verification procedures and
- communication channels to transfer information from the sources to the NPPO.

---

[5] Such as foot and mouth disease, African swine fever, swine vesicular disease etc. Animal Health Australia, & Plant Health Australia (2013). Farm biosecurity. http://www.farmbiosecurity.com.au/. Accessed 15 April 2018..

[6] Note, biosurveillance also refers to the monitoring of the spread of human diseases but the use of the term is restricted in this paper to surveillance for threats to agricultural production.

For example, the Australian Plant Pest Database (APPD) is a national, online database of pests and diseases of Australia's economically important plants used to "assist bids for market access and to justify measures to exclude potentially harmful, exotic organisms, help in emergency plant pest management, and support relevant research activities" ([Plant Health Australia, 2001](#))[7]. Biosurveillance can be either *general* (strategic actions to manage surveillance goals) or *specific* (outlining specific actions of surveillance) according to the *Guidelines on Surveillance* ([Secretariat of the International Plant Protection Convention (IPPC), 2016](#))—see *Table 1.* The guidelines ensure the scientific validity of biosurveillance activities and 'good surveillance practice' means that personnel are scientifically trained with adequate knowledge of plant protection and data management:

> Personnel involved in general surveillance should be adequately trained in appropriate fields of plant protection and data management. Personnel involved in surveys should be adequately trained, and where appropriate audited, in sampling methods, preservation and transportation of samples for identification and record keeping associated with samples. Appropriate equipment and supplies should be used and maintained adequately. The methodology used should be technically valid ([Secretariat of the International Plant Protection Convention (IPPC), 2016, pp. ISPM 6-7](#))

The guidelines suppose that surveyors or 'scouts' must be able to accurately identify the presence and threat level of pests and diseases including how to interpret the severity of an outbreak from limited samples and how to interpret symptoms with multiple potential causal factors, e.g. stunted growth can result from a variety of causes including lack of water, nutrition and varied diseases). Note that there is no discussion of the ethics or best practice to *limit* the collection of information. There is no discussion of how to manage the emergence of autonomous and ubiquitous biosurveillance technologies and enterprises relative to human wellbeing.

*Table 1.* Use of Information for General Surveillance and Specific Surveillance Plans ([Secretariat of the International Plant Protection Convention (IPPC), 2016](#))

---

[7] The APPD draws on the collections and databases from 18 regional databases across Australia.

> *General Surveillance*
> General information gathering will most often be used:
> - To support National Plant Protection Organisations (NPPO) declarations of pest freedom
> - To aid early detection of new pests
> - For reporting to other organisations such as the nine regional protection organizations (RPPOs) and Food and Agriculture Organisation of the United Nations (FAO)
> - In the compilation of host and commodity pest lists and distribution records
>
> *Specific Surveillance*
> Specific surveys may be detection, delimiting or monitoring surveys. These are official surveys and should follow a plan which is approved by the NPPO.
> Survey plans should include:
> - Definition of the purpose (e.g. early detection, assurances for pest free areas, information for a commodity pest list) and the specification of the phytosanitary requirements to be met
> - Identification of the target pest (s)
> - Identification of scope (e.g. geographical area, production system, season)
> - Identification of timing (dates, frequency, duration)
> - In the case of commodity pest lists, the target commodity
> - Indication of the statistical basis (e.g. level of confidence, number of samples, selection and number of sites, frequency of sampling, assumptions
> - Description of survey methodology and quality management including an explanation of:
>   - Sampling procedures (e.g. attractant trapping, whole plant sampling, visual inspection, sample collection and laboratory analysis); the procedure would be determined by the biology of pest and/or purpose of survey
>   - Diagnostic procedures
>   - Reporting procedures

The autonomous collection of data by third parties contrasts with traditional methods of crop biosurveillance that are human-resource-heavy, requiring observation and paper-based recording of plant leaves, stems, seeds, flowers and fruit for insects—usually by limited and trusted groups of growers, agronomists, or scientists—recording pest type and numbers across specific sites—e.g. see pest surveillance data sheet *Figure 2*.

Physical surveillance data records by farmers are usually used to communicate with farm stakeholders and are not necessarily collected by the government or digitized for future examination. However, growers are expected to store data sheets and may be required to work with government to provide evidence of an absence of pests for export and interstate markets.

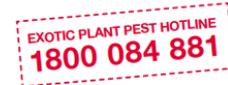

*Figure 2*: A pest surveillance data sheet is printed and used to observe and record the type of pests and quantity across set areas ([Farm Biosecurity Program, 2012](#)). Used with permission from Plant Health Australia.

Reasons to submit a report and/or specimen to a government database emerge if it improves the knowledge of particular scientific taxon ([Plant Pathology Herbarium and Insect Collection Queensland, Beasley and Shivas, 2017](#)) such as:

- New occurrence within the region
- Extension of host range
- Extension of geographical range
- Morphological variation
- Detection of restricted matter or prohibited matter
- Voucher material to support research and scientific publications

Growers are trusted to report unusual finds and to work collaboratively with authorities. Those with good records are able to work with government and industry bodies to prepare emergency chemical registrations, and arrange permits for importation of biocontrol agents for a more rapid and effective response to outbreaks ([See 'Pest Surveillance' in Animal Health Australia and Plant Health Australia, 2013](#)). Good records also help farmers negotiate risk with financial organisations and insurers.

In one sense there is no limit to the amount of information that could be gathered to 'improve knowledge of a particular scientific taxon', but in practical terms, surveillance has been expensive and time-consuming. To access a property requires explicit permission of landholders or managers and appropriate times and dates to be coordinated in advance or in response to invitation. In this way surveillance is a negotiation between land owners and those that wish to surveil. Allowing farm managers or owners to decide when and how surveillance occurs does mean accommodating for human error (whether intentional or accidental) making more systematic and automated surveillance appealing to Government agencies. Already satellite imagery has provided private and public operators much information about the size of farming operations and the nature of the enterprise itself including metrics of crop growth via NDVI. However, satellite monitoring is considered an invasion of privacy to over half of participants when surveyed ([Purdy, 2011, p. 205](#)). Even so, research and development of satellite imagery continues unabated; improving its resolution and accuracy, whilst remaining expensive and intermittent given variability in cloud cover.

The promise of more automated and persistent surveillance via RPAS opens up the possibility of widespread surveillance over farming lands at high resolution with even greater threats to privacy. There are real differences in privacy risk between storing A4 pages on-farm and the potentially vast data-sets kept and accessible in the cloud. There has already been research on violations of privacy and threats to civil liberties through use of RPAS in military, civilian and biodiversity conservation domains (e.g. [Kreps and Kaag, 2012](#); [Sparrow, 2009](#); [Sandbrook, 2015](#)). RPAS allow individuals to monitor other people without their knowledge, raising the question: when does such actions become an unacceptable infringement of privacy or other human rights, such as freedom of association ([Finn and Wright, 2012](#))? RPAS enable farmers to surveil their neighbours, for example to establish the source of spray drift, use of a banned chemical (through high resolution photographs of containers and bins) Governments could conduct checks for unsafe application of chemicals via poor OH&S processes. Farmers could identify pest incursions on their neighbours' properties and report them to authorities for poor farm hygiene. Or they could use RPAS to monitor horticultural techniques and gain a competitive advantage. Note that non-pest and disease related surveillance activities could be completed using data captured by RPAS doing otherwise defensible biosurveillance runs. That is, if individuals or organisations have access to surveillance data, or the ability to acquire it directly, it can be used for much more than looking for pests and diseases. It is important also to note, that the threat of loss of privacy from RPAS is more than just protecting individuals and enterprises from being 'revealed' because they have something to hide. The value of privacy is multifaceted including the right to be free from intrusion in one's life ([Scanlon, 1975](#)) , the right to control and manage information about oneself ([Nissenbaum, 2009](#)) and one's enterprise to facilitate social and business relationships and to protect one's competitive advantage ([Rachels, 1975](#)). The value of privacy pertains to diverse concepts such as knowledge, dignity and freedom ([Post, 2000](#)) and types of privacy along axis of *Freedom* and *Access* ([Koops et al., 2016](#)) discussed more fully in §6.3.

# 5. Surveillance, Privacy and Autonomy

Biosurveillance must be considered against theories of surveillance more generally and particularly the impact of increasing personal and commercial data surveillance from ubiquitous and affordable technologies.

## 5.1.  The history of surveillance

Domestic surveillance has historically been intermittent and targeted. Due to the prohibitive cost of human resourcing and surveillance technologies (e.g. wire tapping), police and government agencies would only surveil citizens they felt had a high likelihood of offending and thus where costs and intervention could be justified. Similarly, traditional biosurveillance ground surveys and ground traps are expensive and time consuming, thus have typically been used sparingly and strategically. Public records built from surveillance have been used to assist governments making policies for the public good. For example, cholera death and locational data helped reveal the source of contaminated water in London, thus the aetiology of the disease ([Cliff et al., 1998](#)). But also for political control, for example, the collection of German census data in 1933 collected on IBM computers enabled Nazis to efficiently target specific ethnic and minority groups ([Black, 2001](#)).

Governments justify domestic biosurveillance for two reasons:
1) because it is likely to return greater economic benefits to the entire community that undergoes it, even if it is more painful or results in loss to some individuals within that community.
2) because it is likely to return greater environmental benefits to the entire ecosystem even if it requires a disproportionate amount of work on some component parts (e.g. incursions on a specific farm or areas with higher identified risk).

Given its advantages to agricultural productivity and food and environmental security, what are the arguments against surveillance? Fundamentally, surveillance is a threat against privacy. Why is privacy important? Perhaps if farmers have nothing to hide, they should not worry about surveillance? Unfortunately, knowing what needs to be 'hidden' may depend on tracking the whims of those in power. Political parties can seek to penalize farmers who have applied specific chemicals or changed their environment against their preferred policies. Farmer livelihoods can be threatened when governments make policy decisions that take autonomy away from rural communities in order to attract urban votes. For example, farmers in Queensland Australia have been accused of excessive land clearing ([The Economist, 2018](#)). But cattle farmers say GIS mapping data draws inaccurate conclusions about their farming practice and the resultant policies take away their right of stewardship over their land in terms of vegetation management ([Queensland Food Future, 2018](#)). Farmers are faced with a dilemma. On the one hand, their social license to operate farming enterprises depends on them being transparent about their practices. But, being transparent could lead to politically selective use of data that is against their interests. On the other hand, farmers who fight to keep their enterprises confidential may lose their social license regardless of whether they are operating sustainably or not.  In sum, governments know they benefit from having more and more information about citizens and farmer autonomy is undermined from a loss of privacy.

## 5.2. Modern surveillance

The defining characteristics of modern surveillance are cheap, usable, ubiquitous, nimble and autonomous. Modern surveillance and A.I. devices are getting cheaper to manufacture and purchase making them easier for smaller groups and private individuals to act without ethical oversight ([Metz, 2018](#); [Brundage et al., 2018](#)). Open source software, refined user interfaces and intuitive apps makes it easier for users to 'just get going' with RPAS without needing any training or certification to consider ethical implications of the technology's use ([Brundage et al., 2018](#)). Modern surveillance devices are also ubiquitous ([Acquisti et al., 2015](#)) with mobile phones, smart watches and fitness trackers capturing intimate personal data 24hrs a day including our physical location, well-being, ideas, desires, plans, questions and interests; our social and professional networks. Secondary personal domestic technologies such as laptops, tablets, IoT devices, personal home-shopping devices etc. build a more comprehensive societal data set. Public spaces are easier for governments to surveil with smaller and more hidden recording devices that they may not adequately warn the public about or give the public meaningful choices over whether they will be recorded. Devices are also nimble, e.g. RPAS are smaller, stronger, with longer battery life and increasingly easy to use for aerial, underground and undersea surveillance. Devices can be placed almost anywhere that a human or government wants to find out more information. Finally, surveillance devices are increasingly autonomous. The algorithms set in motion by human designers enable them to operate for long periods of time without oversight. This combination of cheapness, availability, usefulness and ease of use modern surveillance presents new and significant risks to privacy. Therefore, we believe the freedom to negotiate surveillance is the core right to be protected and that rational persuasion must be undertaken between individuals, government agencies, business and organisations to ensure surveillance behaviours are agreed to by fully informed and willing agents.

## 5.3. CASE STUDY: Autonomous technologies with cameras and machine learning to identify pest and disease incursions

RPAS and autonomous tractors have been automating crop damage assessments and conducting biosecurity surveillance for decision support for almost 30 years ([for early RPAS see Herwitz et al., 2004](); [for summary of autonomous tractors see Mayfield, 2016]()). RPAS combine high spatial resolution, low cost and fast turn around times for imagery. They also circumvent unfavourable sensing conditions such as dense cloud cover that inhibit the results of satellite imagery ([Berni et al., 2009]()) to provide reflectance indices such as NDVI, TCARI/OSAVI and PRInorm ([Gago et al., 2015]()). Autonomous tractors utilise unclassified GPS technologies and a variety of sensors to map fields and vegetation low to the ground. More recently machine learning has been combined with sensor technology to increase the usefulness and usability of collected data, e.g. trained on infrared thermography to detect foot and mouth disease in infected cattle ([Rainwater-Lovett et al., 2009]()), or trained on thousands of crop images to autodetect pests and diseases ([Hamilton et al., 2016](); [Puig et al., 2015]()). RPAS and autonomous tractors are also being developed to carry out the elimination of pests, diseases and weeds through sprays ([Faiçal et al., 2017]()), manual removal ([McCool et al., 2018]()) and even microwaves ([Brodie et al., 2012]()). The precision of the outputs of underlying mathematical models depends on the quantity and quality of data they process—usually the more data, the better.

RPAS and machine learning have been trialled successfully in surveillance contexts including; white grub damage of a sorghum crop ([Puig et al., 2015]()); yellow spot and leaf rust on wheat ([Hamilton et al., 2016]()) and spot form net blotch and leaf rust on barley ([Hamilton et al., 2016]()). Moving from multispectral to hyperspectral cameras has refined the performance of machine learning algorithms in contexts requiring more complex and nuanced visual signatures such as detecting myrtle rust (*Austropuccinia psidii)* on paperbark tea trees in forests ([Sandino et al., 2018](#)) and targeting grape phylloxera (*Daktulospaira vitifoliae*) in vineyards ([Queensland University of Technology, 2018](#)).

Other technologies are also being used to combine visual imagining and machine learning to surveil farms, such as solar powered, low weight robotic tractors that use advanced robot vision algorithms to systematically detect and destroy weeds or selectively spray for pests and diseases ([Bawden et al., 2017](#); [McCool et al., 2018](#)). The autonomous tractor use-case for biosurveillance is potentially less of a threat to farmer privacy as the cameras are pointed straight down into the field, thus are not capturing imagery of the entire farm that might include people and activities. Also, tractors can potentially perceive and make decisions offline (updating Firmware when appropriate), which means farmers could maintain more control over their data[8]. On the flip-side, if tractors depend on APIs to process visual imagery then potential breaches of privacy could occur if the APIs collected certain kinds of metadata, e.g. location data from the tractor when images are taken. If images were uploaded into a cloud-based processing software, then farmers would also need to know how those images were being used, e.g. are they simply being used to train visual processing algorithms, or are maps made of incursions of pests and diseases by governments, insurers or regulators?

The rapid development and uptake of RPAS fills a critical link between broad-scale satellite-based monitoring and local on-ground detection efforts, enabling integrated hierarchical surveillance (Jones et al. 2006), with large increases in efficiency. Governments and other agencies will likely look to more integrated and multi-level approaches to biosurveillance, leading to further concerns about data storage, access and use. Furthermore, there will be a continued need to transform data into relevant and context-dependent information. This requires considerable capacity in appropriate analytical and predictive modelling methods. Compilation of detailed and complex datasets without concomitant development in expertise and stakeholder engagement may lead to an increase in false alarms (false positives) and erroneous identification of threats, undermining confidence in the big-data approach.

At this point in the paper we have considered the ethical underpinning of surveillance from a rights perspective, explained the importance of biosecurity and biosurveillance and outlined some of the latest technologies for conducting biosurveillance that is changing the way surveillance occurs. The reader may wonder what existing regulations exist with regards to using autonomous biosurveillance. In the following section we discuss the regulation of RPAS as the tool most likely to be used by agencies to undertake large-scale biosurveillance and presents the greatest threat to privacy to landholders from unbidden surveillance activities. For more discussion of legal issues for autonomous tractors on private land see Basu et al. ([2018](#)) and Wiseman et al. ([2018](#)).

## 5.4. RPAS regulation

Aerospace regulatory frameworks for RPAS are emerging and adapting globally[9]. Laws are usually focussed on safety rather than security or privacy, such as deconflicting flight paths and banning use during certain events such as fire or police emergencies(Civil Aviation Safety Authority, 2017). RPAS businesses are concerned that new legislation may stifle RPAS innovation (see Uniform Law Commission, 2019; Whynne and Shapiro, 2018, October 25). In regional areas, farmers are concerned about privacy and trespass, particularly that unregulated RPAS are being used by animal activists or environmentalists to disseminate distorted footage of farm activities (Coote, 2018, August 28). Activists believe that their campaign against animal cruelty is warranted against privacy and biosecurity complaints. In response, some governments have developed recommendations to both make animal welfare practices more transparent and to protect landowners against unauthorised filming or surveillance (Select Committee on Landowner Protection from Unauthorised Filming or Surveillance, 2018). Laws can fall behind societal needs due to the speed of technological innovation against the pace of regulatory change.

---

[8] For a detailed review of technology providers and data considerations see Wolfert, Verdouw & Bogaardt (2017).

[9] See the Global Drone Regulations Database https://droneregulations.info/index.html

In the absence of regulatory or legislative obligations ethical guidelines have been set up by some government agencies, e.g. *Drones and Privacy Principles* by the Office of the Information Commissioner Queensland (OICQ) (2018). These principles give practical advice to agencies on reasonable steps to inform individuals who may be under surveillance where personal or commercial information could be collected. This includes provision of a 'collection notice' notice either before or after activities that advises:

- Why information is being collected
- Details of any law that allows or requires the collection and
- Any entity to whom it is the agency's usual practice to give the information

To build trust and cooperation between stakeholders, the OICQ recommends that RPAS surveillance activities are planned with a community engagement strategy and communicated digitally (e.g. via agency website and social media channels), physically (e.g. signage around the area under surveillance, letter box flyers to affected households, banners at the 'launch site') and via media advertisements.

These guidelines go into detail with regards to data storage and access that mimic many of the progressive data policies of the EU's General Data Protection Regulation (GDPR) (European Parliament and Council, 2016). It is likely that more and more government agencies will establish ethical principles as part of the process to implementing regulation. It is in fact this process of moving from ethical considerations to legal frameworks that motivates us to write this paper and lands us in the final section of our paper regarding recommendations for ethical biosurveillance.

# 6. Recommendations for ethical biosurveillance

It is likely that autonomous technologies will continue to develop more effective capabilities based on frequent and comprehensive smart surveillance of a farmer's lands. The risks to farmer autonomy through surveillance justify the creation of an ethical framework for biosurveillance. We propose the following recommendations for ethical biosurveillance:

1. Establish the boundaries of a biosurveillance social contract;
2. Justify surveillance operations for the farmers, researchers, industry, the public and regulators;

3. Give decision makers a reasonable measure of control over their personal and agricultural data; and
4. Choose surveillance methodologies that give the appropriate information.

## 6.1. Establish Boundaries of a Biosurveillance Social Contract

The danger of scientific and technological developments can be that the Government's desire to control nature becomes an ideology promoted by experts and elites, and reduces the urgency for public and democratic discussion around how technologies should be implemented into society ([Jürgen Habermas, 1987](#)) . Physical isolation, low population and lower material assets combine to make rural communities particularly vulnerable to political acts to invade privacy in the name of scientific advantage. For example, efforts to save the Great Barrier Reef have led to the State Government of Queensland to demand private farm management data, collected under a related government-funded program, be turned over to the Government—as farm inputs such as nitrogen have a significant impact on the quality of water in the Reef catchment ([Kroon et al., 2016](#)). Industry lobby group *AgForce* has fought back by deleting over 10yrs of management data to protect stakeholder privacy ([Smee, 2019](#)). It may be that the Government should have more access to farm management data, but we argue that such access is best negotiated rather than mandated to ensure the dignity of growers and a fair public discourse.

The ethical framework we recommend for ethical biosurveillance is via social contract. A social contract, is a theoretical agreement of social rules that individuals adopt to ensure ethical constraints on behaviour. Participants may 'opt in' to a social contract without being obliged to obey it, and may 'opt out' should circumstances warrant it ([Freeman, 2007](#)). To understand these hypothetical individuals, Rawls ([1971/1999](#)) suggested a thought experiment to consider the ideal society from 'behind the veil of ignorance' in 'the original position'. The original position is a hypothetical scenario that allows participants to abstract from their own interests to ensure cooperation in every situation ([Leben, 2019](#)) by creating rules that would be agreed to by anyone in society *before* they knew what role in society they would play. For example, people under a veil of ignorance might consent to some level of biosurveillance to restrict the damages of irresponsible farmers, even though they could not know whether they are an irresponsible or responsible farmer. Even though this approach may sound too abstract to be pragmatic day-to-day, abstraction is actually the tool that allows technology providers to build autonomous systems that optimise and instantiate cooperative behaviours for diverse stakeholders ([See Ch.4 'Contractarianism' in Leben, 2019](#)).

The original position can be modelled with hypothetical scenarios, presenting likelihoods and possible outcomes to help regulators and farmers understand each other's perspectives and experiences through processes of value-centred design and responsible innovation ([Santoni de Sio and Van den Hoven, 2018](#); [Friedman and Kahn Jr., 2003](#); [van den Hoven, 2013](#)). The scenario method is particularly suited when participants are faced with real-world contexts involving high uncertainty and high complexity ([Teece et al., 2016](#); [P.J.H. Schoemaker, 1995](#); [P.J.H. Schoemaker, 1991](#); [Russo et al., 1989](#)). Additionally we recommend negotiating terms should aim for a rhetorically adequate process such as Habermas's ([2008](#)) discourse theory which recommends:
1. No stakeholders capable of having a relevant contribution are excluded.
2. Stakeholders have an equal voice
3. Stakeholders are free to state an honest opinion
4. There are no sources of coercion to the process

The establishment of a social contract via value-centred design and responsible innovation with a rhetorically adequate process will ensure diverse stakeholders in a region affected by planned biosurveillance activities are ethically consulted. Researchers and government agencies seeking to implement extensive, possibly intrusive biosurveillance ought to illustrate a sequence of scenarios where their programs are implemented, and pest and disease eradication and management become significantly more efficient, effective, fairer and easier for managers and affected individuals than if measures had not been taken.

Communicating the scientific, social and economic justifications for biosecurity measures is crucial to building and maintaining stakeholder trust in biosurveillance. Stakeholders must be invited to participate in decision-making around biosecurity models and data. Advocates for autonomous biosurveillance need to explain the impacts of models and data on the public using evidence-based scenarios that allow the public meaningful engagement across a range of surveillance and biosecurity conditions. A framework such as Cairns Goodwin & Wright's ([Cairns et al., 2016](#)) offers a systematic way to incorporate diverse stakeholder responses to scenarios to aid with decision-making. However, many different methods could be utilized to operationalize the boundaries of a biosurveillance social contract.

## 6.2. Justify surveillance operations

Our second recommendation is that each activity within an incursion surveillance program needs to be *justified* scientifically and socially for farmers, researchers, industry, the public and regulators. Justifications are explanations that enable stakeholder understanding and acceptance of their anticipated benefits. A justification might explain why particular data are needed within a specific timeframe, for example to create a useful model for decision making with anticipated benefits. The intervals of surveillance and reporting might affect the choice of different surveillance strategies (see §3.2) or vice versa. Justifications ought to highlight potential advantages to stakeholders of accepting surveillance, such as achieving a Pest Free Area (PFA) designation to reduce the need for fumigation and increase trade efficiency ([Kalaris et al., 2014](#)).

Four types of biosurveillance activities require justification ([Biosurveillance Science and Technology Working Group, 2013, 14 June](#)):
1. Aberration detection;
2. Risk anticipation;
3. Threat identification and characterization; and
4. Information integration, analysis, and sharing.

*Aberration detection* requires surveillance to establish baseline information about normal on-farm and off-farm operations to compare with unusual and potentially risky circumstances. For example, farmers might accept that a year of intense baseline surveillance at the beginning of a 10-year program is justified in order to enable less frequent, yet more targeted biosurveillance for the subsequent nine years.

*Risk anticipation* requires evaluating the likelihood of incursion and spread of diseases and pests. Activities are justified because the earlier a risk is anticipated, the faster and more efficiently an appropriate response can be deployed. Forecasting is greatly enhanced by integrating ongoing real-time data from a variety of data sources including remote sensing (e.g. RPAS) and fixed, distributed autonomous or semi-autonomous surveillance platforms (Biosurveillance Science and Technology Working Group, 2013, 14 June). Privacy issues emerge from any ongoing collection of agricultural data that dovetails with personal data. For example, to establish spatial risk patterns, regulators might request vehicle or phone GPS logs to analyse human movements across farms and regions to prevent or manage incursions. Ongoing, real-time surveillance for risk anticipation needs to be justified via analysis of surveillance sensitivity (e.g. using scenario tree modelling) and incorporation of pre-existing data to establish risk profiles for a region (Kalaris et al., 2014).

*Threat identification and characterization* requires accurately evaluating a pest incursion. Farmers, researchers, industry, the public and regulators need to be sure that a stated pest or disease is a threat and that the magnitude of the threat is accurately presented such that actions can be rationally undertaken. Surveillance activities to validate the identification and characterization of plant pests and diseases will be justified if they significantly improve the accuracy of reports. The costs to growers of false positive and false negative reports should be highlighted before surveillance is undertaken.

*Information integration, analysis and sharing* requires combining data from multiple farms for greater understanding of pest and disease outbreaks. Such activities need to be justified to individuals as part of a greater project of securing their land and their region over long time periods, rather than protecting immediate assets.

## 6.3. Give decision makers a reasonable measure of control over their personal and agricultural data

Our third recommendation is to give farm decision makers a reasonable measure of control over their land use and personal and agricultural data. While given fresh urgency with the volume and resolution of private data collectable, and new systems to analyse and transform this data, this recommendation stems from enduring philosophical and legal discussions relating to the right to privacy. Privacy has been defined variously as 'the right to be let alone' (Warren and Brandeis, 1890; Rubenfeld, 1989); as a cluster of derivative rights (1975) such as rights to own or use one's own property, the right to one's person, or one's right to decide what to do with their body; and as a general right to have a reasonable measure of control over our self-presentation (and that of what is ours) to others (Marmor, 2015). The right to privacy has been categorised as a right to knowledge, dignity and freedom depending on arguments in its favour (Post, 2000). Stakeholders who have privacy concerns should be able to advocate for themselves using this more nuanced conceptions. Some might ask why individuals cannot act freely while being observed? Because, as Bernal (2014) argues "[I]f we want autonomy, if we want freedom, we need privacy to protect it. (p.ix)" We need privacy rights because information is power, and individuals, governments, businesses and organisations who have access to data are more able to manipulate the people the data is about/of (Zuboff, 2019). That is, the more privacy we give up, the more likely it is that our data will be used to curtail our activities rather than laud them or simply allow them to continue. A person's right to privacy is violated when someone manipulates an information environment—without justification—and significantly diminishes a person's ability to control what aspects of themselves they reveal to others. Privacy is not about withholding information, but about the 'contextual integrity' of information sharing (Nissenbaum, 2009). Violations of privacy still occur due to an inappropriate and improper sharing of information. For example, suppose an industry database is designed to capture private information about the prevalence of a particular crop disease in an agricultural region. The organization may have justification to diminish an individual farmer's control over their information due to the need to aggregate accurate information across multiple, contiguous locations to establish strategic and tactical disease management practices. However, such justifications do not consider the right of farm decision makers over their own data. An immediate negative consequence of endorsing the recommendation to allow farmer decision makers reasonable control over their data might be the diminution of data sets leading to more uncertain analyses and less useful recommendations. Reduced trust in models may result in farmers declining further requests to make their data available for biosurveillance activities.

The concept of 'reasonable control' requires a rational, empathetic and cooperative process to occur between farm decision makers and biosurveillance operators. An unreasonable level of control might be for a farmer to share no data from their farm or activities. It is unreasonable because a lack of data prevents proactive and reactive responses to pest and disease management to the detriment of all. Even if a farmer wished to remain informationally isolated for a long period of time, it is possible that their property would suffer an incursion of some threat during their isolation and thus would benefit from others' data relating to the event. A reasonable level of control might be for a farm decision maker to share property and movement data on a sporadic, yet systematic basis, e.g. each farm in a region is scrutinized intensively for one out of five sampling periods so long as a minimum threshold of farms agree to participate. Farmers would not be justified in opting-out of all surveillance activities and would be liable for harms that occurred as a consequence of their recalcitrance.

A typology of privacy (Koops et al., 2016) may be a useful tool to assist biosurveillance operators, farmers and stakeholders to map out the ways that privacy may be threatened and must be managed. Such a typology differentiates privacy on dual axis of *Freedom* and *Access* (see *Figure 3 A typology of privacy* p.566). The *Freedom* axis ranges from 'the right to 'being let alone' (e.g. bodily privacy) to the freedom of self-development (e.g. decisional privacy). The *Access* axis ranges from the personal zone (e.g. intellectual privacy) through to the intimate zone (spatial privacy), semi-private zone (e.g. communicational privacy) and public zone (e.g. proprietary and behavioural privacy). The authors note that informational privacy exists across all axis and intersects with all eight identified types of privacy: bodily, spatial, communicational, proprietary, intellectual, decisional, associational and behavioural.

Taking the literature on privacy into consideration including conceptions and typology of privacy means a contextual approach to privacy must be assumed, where biosurveillance participants renegotiate to opt in to information analysis projects on a case-by-case basis, rather than allowing *carte blanche* access to their data. Individual privacy measures once negotiated could be implemented with the aid of a privacy exchange authority, an intermediary body that helps individuals secure privacy preferences in the face of big data ([Pascalev, 2016](#)).

## 6.4. Choose appropriate biosurveillance methodologies

Once biosurveillance measures are justified (see *recommendation 2*) and individual privacy considered (see *recommendation 3*), the question of appropriateness remains. Our final recommendation is therefore to choose surveillance methodologies that accommodate the first three recommendations to provide *appropriate* information. Appropriate biosurveillance combines the individual privacy preferences of stakeholders with the scientific justification of their use. It should anticipate information gaps. For example, suppose Farmer J has specified a desire to 'be let alone' two out of every three years. Researchers and government agencies seeking to implement effective biosurveillance strategies need to ask: "What are the best methodologies to accommodate J's stated preference?" rather than simply express exasperation. Surveys, models and algorithms can work respectfully around limited information. Methodologies must ensure that *appropriate* information is attained under surveillance, rather than less nuanced approaches such as obtaining as much data as possible given technical capacities, e.g. RPAS use should be driven by the needs of data models, not the amount of ground coverage, flying time or data storage possible.

## 6.5. Discussion

When is surveillance acceptable and when is it not acceptable? Surveillance is acceptable when it is agreed to by stakeholders appreciating contextual justification for the surveillance. Surveillance measures (including the periodicity and duration of surveillance runs) must be negotiated between:
- land owners/managers,
- expert ecologists,
- agronomists,
- relevant government departments[10]
- industry groups and bodies, e.g. research and development corporations,
- local councils and regional organisations,
- neighbours, community members and local land owners/managers.
- members of the public (including activists)

The economic, reputational and time-saving benefits of surveillance must be considered against potential misuse of surveillance information. Data from farms will be most informative over the long-term when systems of data are considered. That is, individual farm data will be aggregated into the data from neighbouring farms. Regional data will be analysed against national and international trends and benchmarks. There will not be a simple answer that resolves the ethical tension inherent in big data collection and use. Different stakeholders are at risk at different levels of analysis. E.g. an industry body might be at risk from National data, where as a local council is at risk through regional information. However, there can be structural interventions that guide and limit data use and 24inimize misuse.

---

[10] Might be the Department of Agriculture, Forestry and Fishing, but might also be Department of Trade and/or the Department of Environment and Science

The outputs of surveillance activities can impact a stakeholder's reputation; either increasing their clout, or diminishing it. The question remains whether such reputational impacts would be fair? Farmers and industries have long experienced impacts on their reputations and business based on outputs, e.g. the quality of their produce or livestock. But, increasingly the *processes* that create the *outputs* are now being 24inimize24ed. Biosecurity practices contribute to a farm's reputation, sometimes formally via industry certification[11] or informally such as community opinion and value-chain relationships. Farm management processes are increasingly transparent to the value chain and sometimes best-practices are financially rewarded. There is a trend towards transparency that advanced biosurveillance practices would contribute to (Jakku et al., 2018). Precautions need to be taken to reduce losses of dignity and liberty for farmers and 24inimize risks for the public and other farmers. We recommend stakeholders use a typology of privacy (Koops et al., 2016) to systematically consider threats to privacy from surveillance. Additionally stakeholders could demand that surveillance companies employ advanced privacy protection measures, such as "differential privacy" that allows data to be used while keeping user anonymity (Abowd, 2017; Dwork and Roth, 2014). Differential privacy has been used by the US Census Bureau, Google and Apple to manage the risks of copious data. Cryptographic methods could potentially be used to model the likelihood of incursions and risks using real historical data without revealing specific land holdings or farmers. Regional stakeholders could weigh in on their own privacy needs; weighing up whether the benefits of geographic identification would be worth privacy risks.

The risk for the public needs to be minimized from inadvertent capture of data from members of the public traversing land that is being surveilled. There are also risks to regional residents from unusual quarantine recommendations arising from extensive data capture. Stakeholders need honest communication with regards to the fidelity of the pest and disease models and how strictly recommendations from these models must be adhered to?

How will increasing biosurveillance make the world more food secure? There are many non-biosurveillance factors contributing to global food insecurity including limited access to finance, hazards of climate change, conflict and violence, gender inequality and small landholders surviving at a subsistence level rather than generating commercial gains (Global Agriculture & Food Security Program, 2018). It is an important and urgent question to what extent different data practices could help to improve food security for these factors—the topic of another paper. However, one aspect of farming that is very difficult to manage with limited finances, climate change, conflict, inequality and subsistence agriculture is incursions of pest and diseases. Identifying and treating pests and diseases requires knowledge and resources.

---

[11] E.g. the Australian Cotton Industry's MyBMP management certification process certifies farms for their business practices including high standards of biosecurity to make farms eligible for global sustainability programs such as the Better Cotton Initiative. See 'MyBmp: The Australian Cotton Industry's Sustainability Standard' via https://cottonaustralia.com.au/cotton-library/publications/brochures

Better biosurveillance data about incursions can reduce costs and improve the productivity of farms. In this paper we have attempted to negotiate the impact of ubiquitous biosurveillance on human rights, particularly the right to privacy. Negotiated amongst stakeholders, biosurveillance is part of a tool set for managing agricultural ecosystems sustainably (Altieri, 2018) and natural resources (Sandino et al., 2018). Alternatives to increasing biosurveillance is increasing the proportion of controlled environment agriculture such as hydroponics, aeroponics, vertical farms, and indoor autonomous farms where the risks of pests and diseases are minimised in food production and farmland can be rehabilitated to a broader ecological function (Despommier, 2013).

## 7. Conclusion

Ideal biosurveillance operations require close to real-time, unimpeded access to crop, pest and disease on-farm data correlated with climate, weather, human movement and trade practices data (Biosurveillance Science and Technology Working Group, 2013, 14 June). Real-time, location-specific data gathering, within a precautionary approach to biosurveillance objectives, implies frequent, ongoing collection of agricultural land data and human behaviour tracking for aberration detection and proactive intervention. These "ideal" biosurveillance operations clearly do not take individual rights to data privacy and control into consideration. We argue that excellent biosurveillance systems can be put in place to meet biosecurity needs while respecting individual rights to privacy on their own land.

In specific we have argued that physical and digital privacy in the age of ubiquitous aerial and ground surveillance is a natural right to allow people to function freely on their land. Surveillance methods must be co-created and justified through using ethically defensible processes such as discourse theory, value-centred design and responsible innovation to forge a cooperative social contract between diverse stakeholders.

Given that laws can fall behind the speed of technological change, it is likely that more and more government agencies will establish ethical principles as part of the process to implementing ethical biosurveillance regulation. These guidelines are likely to mimic many of the progressive data policies of the EU's General Data Protection Regulation (GDPR) (European Parliament and Council, 2016). In this paper we aim to provide policy-makers and regulators clear guidance on how to approach ethical biosurveillance from both a scientific and social perspective. Our four recommendations for ethical biosurveillance are:
1. Establish the boundaries of a biosurveillance social contract;
2. Justify surveillance operations for the farmers, researchers, industry, the public and regulators;
3. Give decision makers a reasonable measure of control over their personal and agricultural data; and
4. Choose surveillance methodologies that give the appropriate information.

Planning and designing biosurveillance within an ethical framework requires moderate changes to data measurement, biosurveillance models, use, access and control of data. Research into biosurveillance innovation ought to include both the positive and negative implications of the technologies proposed (Hecht et al., 2018). The current surveillance guidelines for good practice ought to be updated to ensure that Free Prior and Informed Consent (FPIC) ([United Nations, 2007 Article 10](#)) is obtained by those to be surveilled including the scientific and political justification of proposed surveillance, the potential advantages to the participant (e.g. early detection) and participant's right to deny being surveilled (within reason) without repercussions. If governments wish to enforce an act of biosurveillance, then they must justify both the specific piece of land under scrutiny (e.g. desire to benchmark for 10yrs, or an outbreak has been discovered nearby) and the type of surveillance intended with the participant holding the right of reply to challenge these proposed actions and propose alternate actions. Stakeholders with privacy at risk from data collection and storage should demand proactive data protection such as the implementation of a privacy exchange authority ([Pascalev, 2016](#)) and differential privacy controls ([Dwork and Roth, 2014](#)).

An ethical framework ought to be included in international biosecurity guidelines, such as the United Nations International Plant Protection Convention (IPPC) Guidelines on Surveillance. The benefits of incorporating an ethical framework include resolving the value conflicts and increased adoption by many participants and accumulated trust over time. Long term trust and cooperation will support food security, producing higher quality data overall and mitigating against anticipated information gaps that may emerge due to disrespecting landholder rights.

## 8. Acknowledgements

We are grateful to Angela Daly and Tony Clarke and anonymous reviewers for comments that significantly improved the manuscript. PB and GH acknowledge support of the Australian Government's Cooperative Research Centres Program.